\documentclass[authoryear,preprint,review,12pt]{elsarticle}
\usepackage{amssymb}
\usepackage{amsthm}
\usepackage{natbib}
\usepackage{epsfig}
\journal {New Astronomy}
\begin{document}
\begin{frontmatter}
\title{A CCD photometric study of the late type contact binary EK Comae Berenices}
\author[du] {Sukanta Deb\corref{cor1}}
\ead{sdeb@physics.du.ac.in; sukantodeb@gmail.com}

\author[du] {Harinder P. Singh}

\author[du] {T. R. Seshadri}

\author[iucaa] {Ranjan Gupta}
\cortext[cor1]{Corresponding author. Present Address: Department of Physics \& Astrophysics, University of Delhi, Delhi 110007, India}
\address[du]{ Department of Physics \& Astrophysics, University of Delhi, 
 Delhi 110007, India}
\address[iucaa]{ Inter-University Centre for Astronomy and Astrophysics (IUCAA)
, Post Bag 4, Ganeshkhind, Pune 411007, India}
\begin{abstract}
We present CCD photometric observations of the W UMa type contact binary EK 
Comae Berenices using  the 2 metre telescope of $IUCAA$ Girawali Observatory, 
India. The star was classified as a W UMa type binary of subtype-W by 
\citet{sam1996}. The new V band photometric observations of the star reveal 
that shape of the light curve has changed significantly from the one observed 
by \citet{sam1996}. A detailed analysis of the  light curve obtained from the 
high-precision CCD photometric observations of the star indicates that EK 
Comae Berenices is not a W-type but an A-type totally eclipsing W UMa contact 
binary. The photometric mass ratio is determined to be 0.349 $\pm$ 0.005. A 
temperature difference of $\Delta T = 141 \pm 10 $ K between the components 
and an orbital inclination of $i [^{o}] = 89.800 \pm 0.075$ were obtained for 
the binary system. Absolute values of masses, radii and luminosities are 
estimated by means of the standard mass-luminosity relation for zero age 
main-sequence stars. The star shows O'Connell effect, asymmetries in the light 
curve shape around the primary and secondary maximum. The observed O'Connell 
effect is explained by the presence of a hot spot on the primary component.
\end{abstract}
\end{frontmatter}
\section{Introduction}
The eclipsing binary EK Comae Berenices [hereafter EK Com] was discovered by 
\citet{kin1966}.~They classified the star as a W UMa type binary.~The star was
also observed visually by \citet{bor1990} who determined the period of $0.2666864$ days.~The first photometric solution of EK Com was obtained by \citet
[hereafter SA96]{sam1996} based on the observations of 1994 with the help of 
the Wilson-Devinney [hereafter WD] code. These authors determined a period of 
0.2668726 days and derived other photometric elements from their observed light
 curves and found the evidence for a period decrease at a rate of $\sim 1.5 
\times 10^{-10} \rm cycle^{-1}$. From the light curve solutions using the WD code, they classified the star as a W UMa 
type binary of subtype W with a mass ratio $q = 0.304$, a temperature 
difference of $\Delta T = 310$ K between the components and a fill-out factor 
of 15\%. They estimated the temperatures of the components as $T_{1} = 5000$ K 
and $T_{2} = 5310$ K respectively, where the subscripts 1 and 2 refer to the 
more massive (primary) and the less massive (secondary) component respectively.
 The change in the orbital period of EK Com has been analysed by 
\citet{liz2006} and found the evidence of continuous decrease of orbital period 
$dP/dt = -5.96~\times 10^{-10}$ days cycle$^{-1} = -0.071$ s yr$^{-1}$.  

Since there is no photometric measurement of the star available since 
1994, we have obtained a complete and well-defined light curve of EK Com 
through high-precision CCD photometry using the $IUCAA$ Girawali Observatory 
(IGO) 2-m telescope. The newly obtained light curve is quite different in 
shape than that obtained by SA96. SA96 classified it as a W UMa binary of 
subtype-W by pointing out that the light curve showed an interval of constant 
light in the primary eclipse. However, it is difficult to make out whether a 
minimum is curved or flat and SA96 modelled the light curve assuming a 
flat-bottom primary minimum. Using the new observations, we are able to 
present highly accurate and more precise light curve of EK Com than SA96. 
Based on the appearance of the eclipses in our data, we conclude that the 
system is clearly an A-type W UMa binary, i.e, the deeper eclipse is the 
transit of the more massive star by the less massive one. Our new photometric 
observations, in fact, clearly suggest that the primary minimum is rounded and 
the secondary minimum is flat indicating its A-type nature. 

In this paper, we report the complete V band light curve analysis of the star
EK Com from new photometric observations and employ the WD light curve 
modelling technique.~This technique is used extensively to characterize the 
observed photometric light curves of eclipsing binary stars with the computed 
synthetic light curves thus yielding the parameters of components of the 
binary system. It has been seen from the literature that determination of 
photometric mass ratio $q$ obtained from the WD light curve modelling 
is reliable only for a total-eclipse configuration of the binary system as
 such light curves show characteristic inner contacts with duration of 
totality setting a strong constraint on the $(q, i)$ pair \citep{ruc2001}. 
Since EK Com is found to be a total-eclipsing over-contact binary, the various 
geometrical and physical parameters obtained from the light curve modelling 
are expected to be reliable. Although we have only two nights of data observed 
only in the V band, the phase coverage is excellent enough to reveal the 
accurate shapes of the primary and secondary minima. In Section 2, we present 
the observations and data reduction procedures. Section 3 deals with the 
period analysis and determination of ephemeris from new observations. In 
Section 4, we describe detailed photometric analysis of the star in the V band 
using the WD code. In Section 5, we describe determination of distance to EK 
Com. Finally, in Section 6, we summarise the results and main conclusions of
our study. 
\section{Observations and data reduction}
All the V band CCD photometric observations of the star were carried out 
with the 2-m IGO telescope, located about 80 km from Pune, India during two 
nights on March 30 and April $01$ in 2009. The IUCAA Faint Object 
Spectrograph Camera (IFOSC) equipped with EEV 2 K~$\times$ 2 K thinned, 
back-illuminated CCD with 13.5 $\mu m$ pixels was used. The CCD used for 
imaging provides an effective field of view of $10.5^{'} \times 
10.5^{'} $ on the sky corresponding to a plate scale of 0.3 arcsec 
pixel$^{-1}$. The gain and read out noise of the CCD camera are 1.5 e$^{-}$/ADU
 and 4 e$^{-}$ respectively. The FWHM of the stellar image varied from 3 to 
5 pixels during the observations. We took a total of 296 frames in the V band 
with the exposure times varied between 10s to 60s for a good photometric 
accuracy. 

\begin{figure}
\begin{center}
\includegraphics[height=9cm,width=9cm]{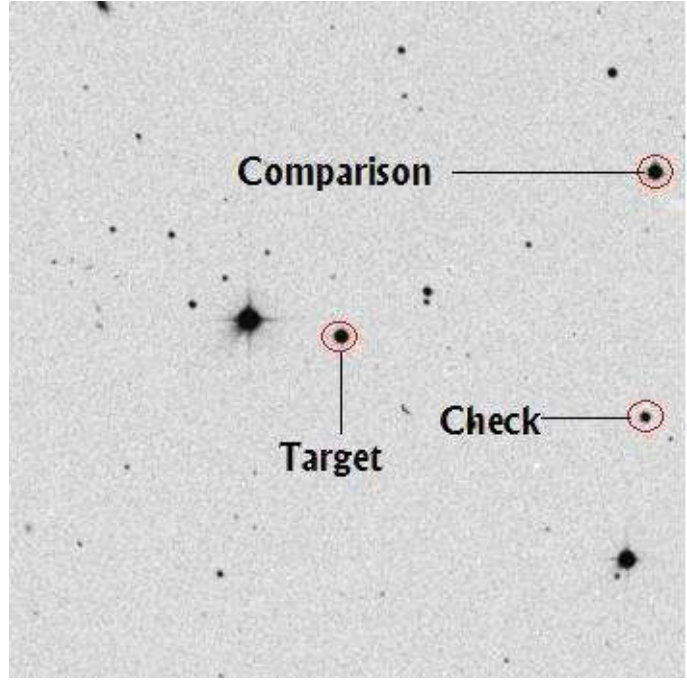}
\caption{A $\sim 10^{'} \times 10^{'}$ image of the field with EK Com
at centre taken from ESO Online Digitized Sky Survey. The target star, 
comparison star and the check star are marked. The range of RA along the 
horizontal axis is ($\rm 12^{h}~51^m~43^{s}.40,~12^{h}~50^{m}~58^{s}.57$) and 
that of DEC along the vertical axis is (\rm $27^{\rm o}~08^{'}~42^{''}.20,~27^{\rm o}~18^{'}~40^{''}.80$).}
\label{image_ekcom}
\end{center}
\end{figure}

The co-ordinates of the variable, comparison star and the check star are 
listed in Table~\ref{table1}. The comparison and the check stars are close to 
the variable and they are in the same field of view during the observations. 
The comparison star used here is the same used as a check star by SA96. 
The standard $V$-magnitude of the comparison star is 12.06 $\pm$ 0.01 (SA96). 
In Fig.~\ref{image_ekcom}, we show a  $\sim10^{'} \times 10^{'}$ image of the field with EK Com at 
centre taken from ESO Online Digitized Sky Survey.

\begin{figure}
\begin{center}
\includegraphics[height=9cm,width=9cm]{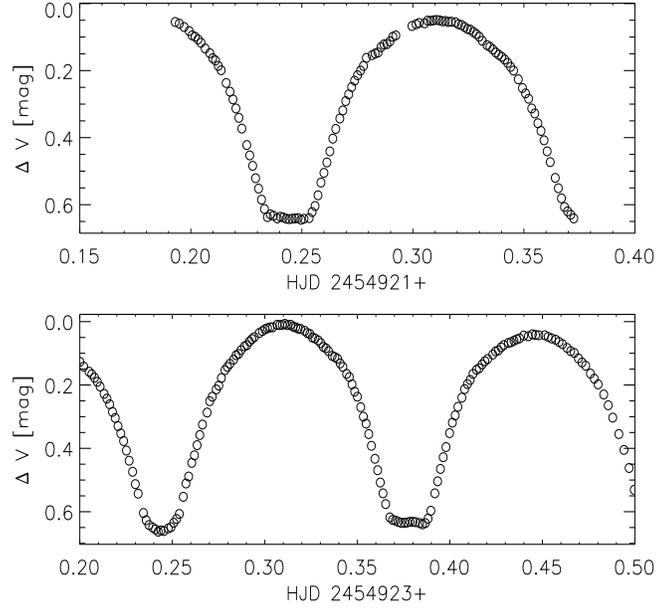}
\caption{Upper and lower panels show the differential magnitudes versus
time (in HJD) of EK Com with respect to the comparison star for observations 
on March 30, 2009 and April 01, 2009 respectively.}
\label{mag_ekcom}
\end{center}
\end{figure}

\begin{figure}
\begin{center}
\includegraphics[height=9cm,width=9cm]{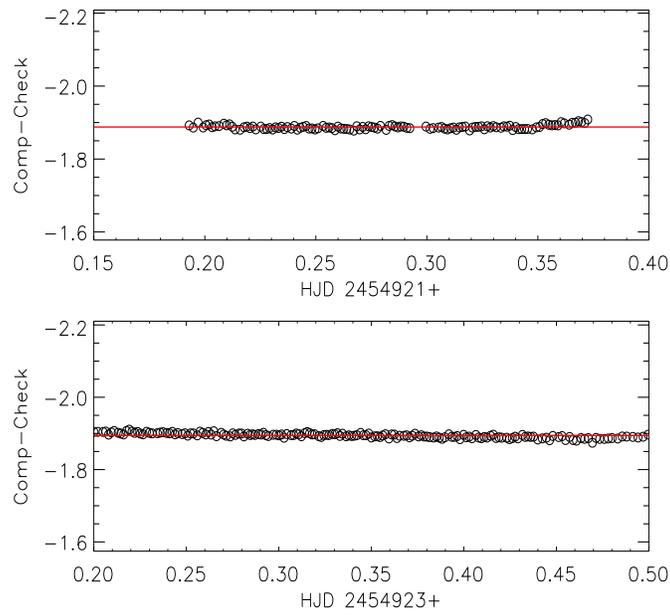}
\caption{Upper and lower panels show the magnitude differences between the 
comparison and check star versus time (in HJD) for observations on March 30, 
2009 and April 01, 2009 respectively. A solid straight line is drawn at the 
mean of the differential magnitude data points to show the constancy of 
variation between the comparison and check star during the course of 
observations.}
\label{comp_check}
\end{center}
\end{figure}

Several bias frames and twilight flatfield frames were taken with the CCD 
camera to calibrate the images of the stars using standard techniques.~Data 
reduction was carried out using IRAF \footnote
{IRAF is distributed by the National Optical Astronomy Observatories, which are operated by the Association of Universities for Research in Astronomy, Inc., 
under cooperative agreement with the National Science Foundation.}
 and MIDAS softwares.~The raw images were processed following the standard 
procedures to remove the bias, trim the images and divide by the mean sky 
flatfield. The resulting images are therefore free from the instrumental 
effects. Instrumental magnitudes were obtained using the {\it DAOPHOT} package 
\citep{ste1987, ste1992}. The various tasks, e.g., {\it find, phot, daogrow, 
daomatch} and {\it daomaster} were applied in order to obtain the instrumental 
magnitudes of stars in all the frames. Extinction corrections were ignored as 
the target star is very close to the comparison star.  In Fig.~\ref{mag_ekcom},
 we show the plots of the V band magnitudes of (Variable $-$ Comparison) versus 
Heliocentric Julian Day (HJD) in the upper and lower panels respectively for 
observations on March 30 and April 01, 2009 and in Fig.~\ref{comp_check}, 
we plot the magnitude differences between the comparison and check star 
(Comp $-$ Check) versus HJD in the upper and lower panels respectively for 
observations on the respective days.

The reduced results show that the difference between the magnitude of 
the check star and that of the comparison star was constant with a probable 
error of $\pm$0.006 mag in the V band. In Table~\ref{table2}, we list the time 
of observations in HJD and the differential magnitudes of the star with 
respect to the comparison star ($\rm \Delta V$).
\begin{table}
\begin{center}
\caption{Coordinates and V band magnitudes of the variable star EK Com, 
comparison star and check star}
\label{table1}
\scalebox{0.9}{
\begin{tabular}{lccc}
\hline
Star  & RA (J2000) & DEC (J2000) & V mag \\ \hline
EK Com & $12^{\rm h}\,51^{\rm m}\,21^{\rm s}.45$ & $27^{\rm o}\,13^{\rm '}\, 47^{\rm ''}.03$ &12.021   \\
Comparison & $12^{\rm h}\,51^{\rm m}\,01^{\rm s}.10$ & $27^{\rm o}\,16^{\rm '} 19^{\rm ''}.70$ &12.060 \\ 
Check & $12^{\rm h}\,51^{\rm m}\,01^{\rm s}.40$ & $27^{\rm o}\,12^{\rm '}\,43^{\rm ''}.00$ &13.998\\ \hline
\end{tabular}
}
\end{center}
\end{table}
\begin{table*}
\caption{V Band CCD observations of EK Com}
\label{table2}
\scalebox{0.5}{
\begin{tabular}{lcccccccccccc}
\hline \hline  
HJD  &  &HJD & & HJD & & HJD & &HJD & &  HJD & &\\
(2,454,000 +)&$\rm \Delta V$ &(2,454,000 + )&$\rm \Delta V$ &(2,454,000 +)&$\rm \Delta V$ &(2,454,000 +)&$\rm \Delta V$&(2,454,000 +)&$\rm \Delta V$&(2,454,000 +)&$\rm \Delta V$\\ \hline
   921.1929&	 0.055&    921.2671&	 0.343&    921.3440&	 0.189&    923.2508&	 0.635&923.3288&     0.059&   923.4080&     0.213& \\	      
   921.1948&	 0.060&    921.2684&	 0.319&    921.3453&	 0.200&    923.2523&	 0.623&923.3302&     0.068&   923.4095&     0.197& \\	      
   921.1970&	 0.070&    921.2698&	 0.291&    921.3475&	 0.226&    923.2538&	 0.607&923.3316&     0.080&   923.4110&     0.184& \\	      
   921.1993&	 0.082&    921.2711&	 0.271&    921.3494&	 0.252&    923.2559&	 0.553&923.3330&     0.084&   923.4125&     0.165& \\	      
   921.2006&	 0.095&    921.2724&	 0.249&    921.3507&	 0.268&    923.2574&	 0.511&923.3344&     0.093&   923.4144&     0.152& \\	      
   921.2019&	 0.099&    921.2738&	 0.230&    921.3521&	 0.284&    923.2588&	 0.489&923.3364&     0.108&   923.4159&     0.147& \\	      
   921.2032&	 0.107&    921.2751&	 0.214&    921.3535&	 0.313&    923.2605&	 0.446&923.3378&     0.112&   923.4174&     0.135& \\	      
   921.2046&	 0.117&    921.2764&	 0.200&    921.3549&	 0.328&    923.2620&	 0.422&923.3393&     0.117&   923.4189&     0.125& \\	      
   921.2064&	 0.134&    921.2777&	 0.186&    921.3563&	 0.357&    923.2634&	 0.390&923.3407&     0.132&   923.4204&     0.118& \\	      
   921.2086&	 0.148&    921.2791&	 0.162&    921.3577&	 0.380&    923.2648&	 0.358&923.3422&     0.143&   923.4232&     0.104& \\	      
   921.2098&	 0.163&    921.2817&	 0.154&    921.3591&	 0.408&    923.2663&	 0.325&923.3442&     0.165&   923.4247&     0.094& \\	      
   921.2111&	 0.170&    921.2830&	 0.149&    921.3605&	 0.441&    923.2688&	 0.287&923.3457&     0.175&   923.4262&     0.091& \\	      
   921.2124&	 0.186&    921.2843&	 0.146&    921.3619&	 0.473&    923.2702&	 0.251&923.3471&     0.198&   923.4277&     0.081& \\	      
   921.2137&	 0.199&    921.2856&	 0.130&    921.3641&	 0.520&    923.2716&	 0.237&923.3486&     0.221&   923.4292&     0.074& \\	      
   921.2159&	 0.237&    921.2870&	 0.122&    921.3655&	 0.551&    923.2737&	 0.214&923.3500&     0.237&   923.4316&     0.067& \\	      
   921.2176&	 0.263&    921.2883&	 0.121&    921.3669&	 0.581&    923.2751&	 0.202&923.3518&     0.270&   923.4331&     0.065& \\	      
   921.2190&	 0.286&    921.2896&	 0.113&    921.3683&	 0.607&    923.2765&	 0.178&923.3533&     0.300&   923.4346&     0.060& \\	      
   921.2203&	 0.313&    921.2910&	 0.099&    921.3697&	 0.620&    923.2784&	 0.153&923.3547&     0.322&   923.4361&     0.056& \\	      
   921.2216&	 0.341&    921.2923&	 0.095&    921.3711&	 0.630&    923.2798&	 0.143&923.3562&     0.355&   923.4376&     0.053& \\  
   921.2230&	 0.373&    921.2996&	 0.067&    921.3725&	 0.641&    923.2811&	 0.133&923.3576&     0.392&   923.4399&     0.044& \\	      
   921.2252&	 0.422&    921.3015&	 0.062&    923.2001&	 0.127&    923.2831&	 0.120&923.3595&     0.433&   923.4424&     0.046& \\	      
   921.2265&	 0.453&    921.3028&	 0.058&    923.2021&	 0.142&    923.2845&	 0.106&923.3610&     0.469&   923.4446&     0.040& \\	      
   921.2278&	 0.484&    921.3053&	 0.059&    923.2050&	 0.158&    923.2858&	 0.102&923.3624&     0.508&   923.4463&     0.042& \\	      
   921.2292&	 0.521&    921.3067&	 0.051&    923.2065&	 0.166&    923.2871&	 0.090&923.3638&     0.542&   923.4481&     0.043& \\	      
   921.2305&	 0.552&    921.3080&	 0.052&    923.2080&	 0.177&    923.2890&	 0.078&923.3653&     0.576&   923.4513&     0.043& \\	      
   921.2319&	 0.585&    921.3093&	 0.050&    923.2095&	 0.190&    923.2903&	 0.074&923.3676&     0.618&   923.4531&     0.048& \\	      
   921.2332&	 0.613&    921.3106&	 0.049&    923.2110&	 0.205&    923.2917&	 0.063&923.3691&     0.626&   923.4550&     0.053& \\	      
   921.2345&	 0.637&    921.3120&	 0.050&    923.2131&	 0.228&    923.2930&	 0.055&923.3706&     0.628&   923.4571&     0.057& \\	      
   921.2358&	 0.629&    921.3133&	 0.053&    923.2146&	 0.243&    923.2943&	 0.049&923.3721&     0.632&   923.4590&     0.064& \\	      
   921.2372&	 0.633&    921.3146&	 0.052&    923.2161&	 0.261&    923.2969&	 0.036&923.3736&     0.635&   923.4608&     0.071& \\	      
   921.2390&	 0.641&    921.3160&	 0.055&    923.2176&	 0.285&    923.2987&	 0.030&923.3755&     0.635&   923.4633&     0.083& \\	      
   921.2403&	 0.636&    921.3173&	 0.054&    923.2191&	 0.304&    923.3001&	 0.024&923.3770&     0.634&   923.4653&     0.093& \\	      
   921.2416&	 0.638&    921.3193&	 0.056&    923.2206&	 0.330&    923.3014&	 0.021&923.3785&     0.632&   923.4672&     0.099& \\	      
   921.2430&	 0.642&    921.3206&	 0.062&    923.2221&	 0.353&    923.3028&	 0.018&923.3800&     0.632&   923.4695&     0.116& \\	      
   921.2443&	 0.643&    921.3219&	 0.064&    923.2237&	 0.377&    923.3042&	 0.018&923.3815&     0.633&   923.4714&     0.129& \\	      
   921.2456&	 0.642&    921.3232&	 0.070&    923.2252&	 0.407&    923.3066&	 0.010&923.3837&     0.636&   923.4734&     0.139& \\	      
   921.2470&	 0.641&    921.3246&	 0.076&    923.2267&	 0.439&    923.3080&	 0.011&923.3852&     0.640&   923.4759&     0.164& \\	      
   921.2483&	 0.640&    921.3259&	 0.078&    923.2285&	 0.474&    923.3094&	 0.010&923.3867&     0.636&   923.4780&     0.180& \\	      
   921.2496&	 0.645&    921.3272&	 0.085&    923.2300&	 0.513&    923.3108&	 0.007&923.3882&     0.622&   923.4801&     0.198& \\	      
   921.2509&	 0.642&    921.3286&	 0.091&    923.2315&	 0.543&    923.3122&	 0.011&923.3897&     0.597&   923.4832&     0.230& \\	      
   921.2533&	 0.640&    921.3299&	 0.099&    923.2344&	 0.604&    923.3140&	 0.010&923.3919&     0.542&   923.4856&     0.264& \\	      
   921.2546&	 0.622&    921.3312&	 0.111&    923.2359&	 0.627&    923.3153&	 0.016&923.3934&     0.504&   923.4880&     0.303& \\	      
   921.2560&	 0.604&    921.3334&	 0.124&    923.2374&	 0.642&    923.3166&	 0.018&923.3949&     0.465&   923.4913&     0.355& \\	      
   921.2573&	 0.572&    921.3347&	 0.129&    923.2394&	 0.650&    923.3180&	 0.022&923.3964&     0.428&   923.4941&     0.405& \\	      
   921.2586&	 0.534&    921.3360&	 0.137&    923.2409&	 0.656&    923.3193&	 0.023&923.3979&     0.396&   923.4969&     0.462& \\	      
   921.2599&	 0.505&    921.3374&	 0.146&    923.2424&	 0.663&    923.3215&	 0.029&923.3998&     0.352&   923.4997&     0.531& \\	      
   921.2613&	 0.474&    921.3387&	 0.153&    923.2439&	 0.659&    923.3229&	 0.036&923.4014&     0.319&           & 	 & \\	      
   921.2626&	 0.441&    921.3400&	 0.159&    923.2454&	 0.661&    923.3243&	 0.039&923.4029&     0.296&           & 	 & \\	
   921.2639&	 0.403&    921.3414&	 0.167&    923.2478&	 0.653&    923.3257&	 0.050&923.4044&     0.269&           & 	 & \\	
   921.2653&	 0.374&    921.3427&	 0.177&    923.2493&	 0.649&    923.3271&	 0.053&923.4065&     0.240&           & 	 & \\ \hline 	  	      
\end{tabular}}
\end{table*}	      

\section{Period analysis and determination of ephemeris of EK Com}
\begin{figure}
\begin{center}
\includegraphics[height=9cm,width=9cm]{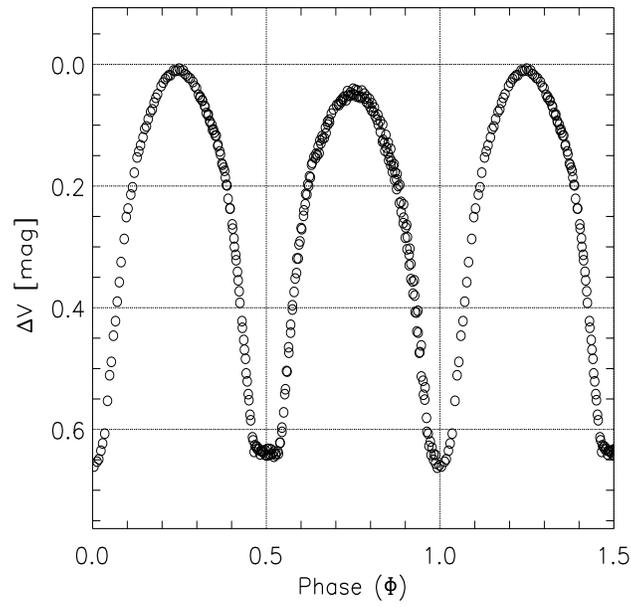}
\caption{Phased light curve of EK Com. The light curve is plotted in the 
phase range [0, 1.5] for a better view of the two minima. Open circles denote 
the observational data points.}
\label{phase_lc}
\end{center}
\end{figure}

A period analysis was carried out using multiharmonic ANOVA algorithm 
developed by \citet[hereafter SC96]{sch1996} to find out the period of EK Com. 
The method computes periodogram by fitting multiharmonic Fourier series to the 
time series data. It is very efficient and robust for non-sinusoidal signals. 
The various statistical properties and the advantages of this method are 
described in SC96. The following times of minima were determined from the data 
using the method given by \citet{kwe1956}:
\begin{displaymath}
\rm Pri.~\rm minima = 2454923.24382(7) 
\end{displaymath}
 
\begin{displaymath}
\rm Sec.~\rm minima = 2454921.24434(6),2454923.37792(5) 
\end{displaymath}
The numbers given in parentheses represent the probable errors. We use the 
following ephemeris to derive the phased light curve.
\begin{eqnarray}
\rm Min I = \rm HJD\,2454923.24382(7) + 0^{d}.26674 \times E,
\end{eqnarray}

where E is the epoch in days. With 2454923.24382 as the time of the primary 
eclipse, we plot the phased light curve of the star in Fig.~\ref{phase_lc} 
which is defined as an array of phase ($\Phi$) and differential V band 
magnitude ($\rm \Delta V$).~The term phase ($\Phi$) is defined as :
\begin{equation}
\Phi =\frac{\left( t-t_{0}\right) }{P}-Int\left( \frac{ t-t_{0}
}{P}\right).
\end{equation}
The value of $\Phi$ is from 0 to 1, corresponding to a full cycle of
the orbital period and $Int$ denotes the integer part of the quantity. The
zero point of the phase corresponds to the time of the primary eclipse 
($t_{0}$).

\section {Photometric analysis using the WD code}

The photometric analysis of EK Com was done by the WD code as implemented in 
the software PHOEBE \citep{prs2005}.~It is a modified package of the 
widely used WD program for eclipsing binary stars 
\citep{wil1971,wil1979,wil1990}.

The appearance of  the eclipses in our data shows that EK Com is likely an 
A-type rather than a W-type system, implying that the deeper eclipse is the 
transit of the massive star by the less massive one. From Fig.~\ref{phase_lc}, 
it is clearly seen that the V-band light curve presents a flatter bottom secondary 
eclipse covering approximately $0.1$ in phase; this possibly indicates a total-eclipse 
configuration of the system. The flatter secondary eclipse is obviously shallower 
than the rounded primary one. It implies that the system is more likely an A-type W UMa 
binary. It seems that EK Com has evolved from W to A type and resembles AH Cnc
which went from W to A type in a time less than a decade 
\citep{san2003,zha2005}.   

We define the massive primary component as star~1 and the less massive 
component as star~2 in the analysis that follows. The effective temperature of 
the primary component can be calculated from the period-color relation given 
by \citet{ruc2001} who derived the following period-color relation for contact 
binary systems.
\begin{eqnarray}
(B-V) = 0.04\,P^{-2.25},
\end{eqnarray}  

where $P$ is the period in days.~With the above equation, the color of 
EK Com can be calculated as: $(B-V)$ = 0.782. The interstellar extinction along 
the direction of EK Com is $E(B-V) = 0.010$ following \citet{schl1998}. 
Therefore the intrinsic color index of the star would be $(B-V)_{0} = 0.772$.
On the other hand, the infrared color index for the star is $(J-K) = 0.560$ 
following \citet{cut2003}. Both these color indices suggest a spectral type 
nearly $\rm K{0}V$ for the binary system.

We take the temperature of the  star~1  $T_{1} = 5150~K$ suitable for the 
spectral type $\rm K0V$ through the calibration of \citet{cox2000}.~The 
gravity-darkening exponents were taken to be $g_{1} = g_{2} = 0.32$ 
according to \citet{luc1967} and albedos adopted as $A_{1} = 
A_{2} = 0.5$ following \citet{ruc1969}.~The monochromatic $(x_{1V}, x_{2V}, y_{1V}, 
y_{2V})$ and bolometric $(x_{1bol}, y_{1bol}, x_{2bol}, y_{2bol})$ 
limb-darkening coefficients of the components were interpolated for square 
root law from \citet{van1993} tables.  The adjustable parameters are the 
mass ratio $q$, orbital inclination angle $i$, the mean temperature of the 
secondary $T_{2}$, the surface potentials of the components $\Omega_{1}$ and 
$\Omega_{2}$, and the monochromatic luminosity of the star~1 $L_{1}$ (the 
planck function was used to compute the luminosity).

\begin{figure}
\begin{center}
\includegraphics[height=9cm,width=9cm]{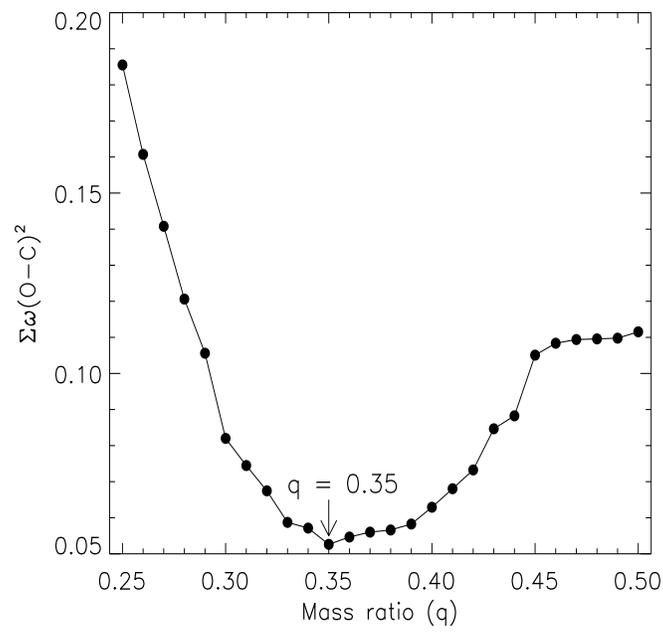}
\caption{Sum of the square of the residuals $\Sigma \omega(O-C)^2$ as a function of the mass ratio ($q$). The minimum of $\Sigma \omega(O-C)^2$ occurs at $q = 0.35$.
}
\label{residual}
\end{center}
\end{figure}

\begin{figure}
\begin{center}
\includegraphics[height=9cm,width=9cm]{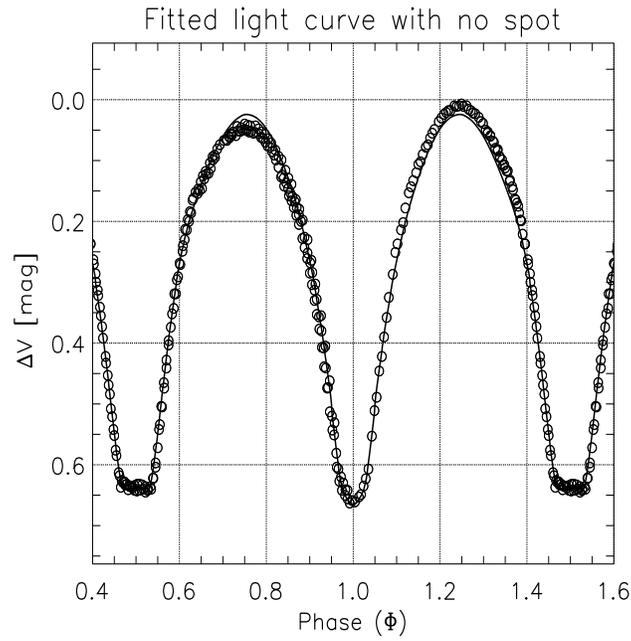}
\caption{Solid line shows the synthetic light curve computed
with the parameters given in Table~\ref{table3} with no `spot' considered on the
components. The weighted sum of the squared residuals of the fit parameters is $\sum \omega_{i} {(O-C)_{i}}^2$ = 0.055}
\label{sol_nospot}
\end{center}
\end{figure}

To search for a reliable mass ratio $q$, we made test solutions at the outset. The test solutions were 
computed at a series of assumed mass ratios $q$, with the values from $0.25$ 
to $0.5$ in steps of 0.1, although SA96 quoted a $q = 0.304$ for EK Com. 
Assuming that it is a detached system, the differential corrections were 
started from the mode 2 and rapidly ran into mode 3 (contact). After several 
iterations, a convergent solution was obtained for each assumed value of $q$. 
From Fig.~\ref{residual}, it can be seen that resulting sum of the squared 
deviations $\sum \omega_{i} (O-C)^{2}$ of the convergent solutions for each 
value of $q$ show that the fit is best in the case of $q = 0.35$. Starting 
from the DC solutions at $q = 0.35$, we ran the code again and let the mass 
ratio be adjusted freely along with the other adjustable parameters. The mass 
ratio converged to a value of $q =0.349$ in the final solution. This value of 
$q$ indicates that EK Com is a A-subtype W UMa binary. With $q = 0.349$, the 
final solution of the system was obtained.  The derived photometric parameters 
are listed in Table~\ref{table3}. The radii $r$ given in Table~\ref{table3} are 
normalised to the semi-major axis of the binary system, i.e., $r= R/a$. The 
component radii are taken as the geometrical mean of the polar, side and 
back radii. Fig.~\ref{sol_nospot} shows the theoretical light curve 
computed with these parameters. 

\begin{figure}
\begin{center}
\includegraphics[height=9cm,width=9cm]{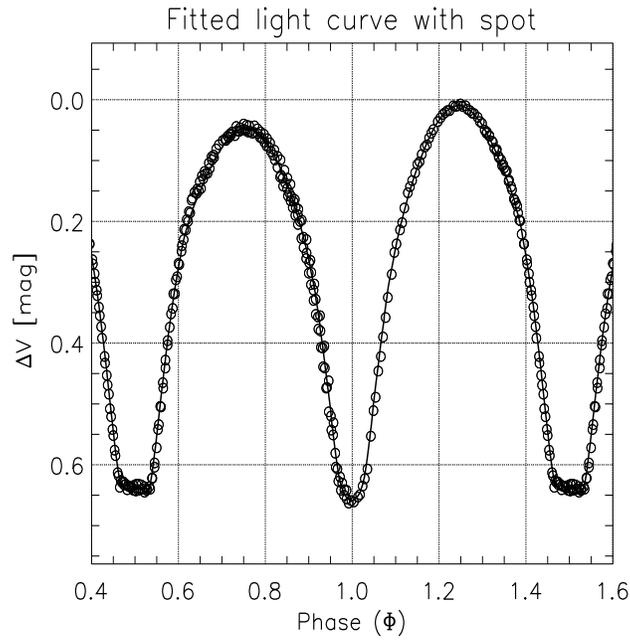}
\caption{Solid line shows the synthetic light curve computed
with the parameters given in Table~\ref{table3} with `spot' considered on star 1. 
The weighted sum of the squared residuals of the fit parameters is $\sum \omega_{i} {(O-C)_{i}}^2$ = 0.016.}
\label{sol_spot}
\end{center}
\end{figure}

\begin{figure}
\begin{center}
\includegraphics[height=9cm,width=9cm]{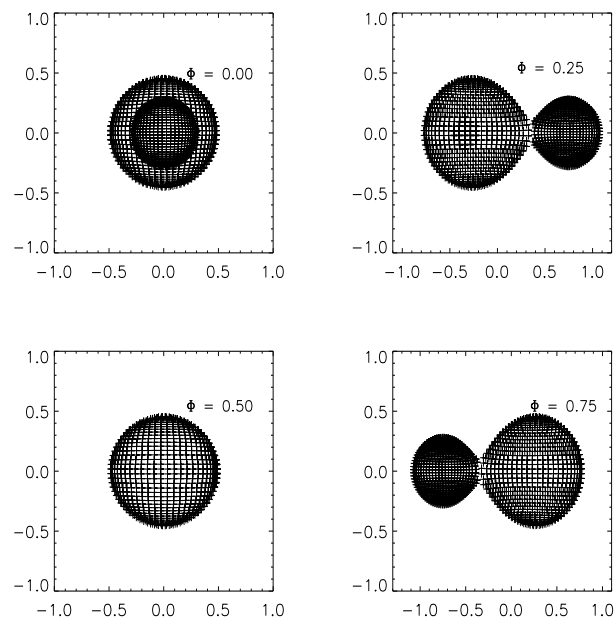}
\caption{Configurations of EK Com in phases of 0.00, 0.25, 0.50 and 0.75 with 
a hot spot on the primary component.}
\label{conf}
\end{center}
\end{figure}

It is very clear that the theoretical light curve does not fit the observational 
data very well, especially around the quadrature levels. The observed light curves 
are asymmetric, with the primary maximum brighter than the secondary maximum. 
This kind of phenomenon of unequal heights at quadrature levels, called \citet{oco1951} 
effect exists in many binaries. We assumed a hot or a cool spot located on either 
the primary or secondary component. We found convergent and stable solutions with 
a hot spot on the primary component. Spot parameters, listed in 
Table~\ref{table3} are co-latitude  $\phi$, longitude $\theta$, angular 
diameter $r_{s}$, and the temperature factor $T_{s}/T_{\star}$ ($T_{s}$ is the 
temperature of the spot, and $T_{\star}$ is the local effective temperature of 
the surrounding photosphere). The solid line in Fig.~\ref{sol_spot} shows the 
fit to the light curve considering a hot spot on the primary component. The 
corresponding configurations of EK Com with a hot spot on the primary component
 are plotted in Fig~\ref{conf}.

The fill-out factor or degree of over-contact ($f$) is given by 
\begin{eqnarray}
f = \frac{\Omega_{\rm in} - {\Omega}}{\Omega_{\rm in} - \Omega_{\rm out}}, 
\end{eqnarray}
where $\Omega_{\rm in}$ and $\Omega_{\rm out}$ refer to the inner and the 
outer Lagrangian surface potentials respectively and $\Omega$ represents the 
dimensionless potential of the surface of the common envelope for the system. 
The degree of over-contact was determined to be 33.0\%.
\begin{table}
\caption{Photometric solutions of the Eclipsing binary EK Com}
\label{table3}
\scalebox{0.9}{
\begin{tabular}{ccc}
\\
\hline \hline 
Parameters &with `no spot'& With `spot' on star 1 \\
\hline
$q $ & 0.349 $\pm$ 0.005 & 0.346 $\pm$ 0.005 \\
$i [^{o}]$ & 89.800 $\pm$ 0.075 & 89.808$\pm$ 0.085 \\
$L_{1}/(L_{1} + L_{2})$ & 0.689 $\pm$ 0.002 & 0.689 $\pm$ 0.002 \\
$x_{1V}$ = $x_{2V}$  & 0.526 & 0.526 \\
$y_{1V}$ = $y_{2V}$  & 0.256 & 0.256 \\
$x_{1bol}$ = $x_{2bol}$  & 0.218 & 0.218 \\
$y_{1bol}$ = $y_{2bol}$  & 0.452 & 0.452 \\
$\Omega_{1} = \Omega_{2}$ & 2.501 $\pm$ 0.003 & 2.521 $\pm$ 0.003 \\
$\Omega_{\rm in}$  & 2.572 & 2.566\\
$\Omega_{\rm out}$ & 2.357 & 2.352 \\
$T_{1}$ & 5150 K & 5150 K \\
$T_{2}$ & 5291 $\pm$ 10 K & 5307 $\pm$ 5 K \\
$A_{1} = A_{2}$ & 0.500 & 0.500 \\
$g_{1} = g_{2}$ & 0.320 & 0.320 \\   
$\sum$ & 0.054& 0.016 \\
$r$(pole): & & \\
star 1 & 0.4568 $\pm$ 0.0006 & 0.4549 $ \pm$ 0.0005  \\
star 2 & 0.2861 $\pm$ 0.0007 & 0.2826 $ \pm$ 0.0005 \\
$r$(side): & & \\
star 1 & 0.4926 $\pm$ 0.0009 & 0.4899 $\pm$ 0.0007 \\
star 2 & 0.3002 $\pm$ 0.0009 & 0.2960 $\pm$ 0.0007 \\
$r$(back): & & \\
star 1 & 0.5236 $\pm$ 0.0012 & 0.5199 $\pm$ 0.0009\\
star 2 & 0.3446 $\pm$ 0.0016 & 0.3377 $\pm$ 0.0012\\
Spot: & & \\
$\phi[^{\rm o}]$ & $...................$ & 90.000 \\
$\theta[^{\rm o}]$ & $...................$ & 260.933 \\
$r_{s}[^{\rm o}]$ & $....................$ & 16.000 \\
$T_{s}/T_{\star}$& $....................$& 1.097 \\
\hline
\hline
\end{tabular}
}
\end{table}


Since there is no spectroscopic measurement of the orbital elements available 
presently, the absolute parameters of the system cannot be determined directly.
We use the \citet[hereafter MV96]{mac1996} method to derive the total mass and 
total luminosity of EK Com.~Then using the mass ratio determined from the 
photometry, we can derive the individual masses, luminosities etc. The 
absolute dimensions and the luminosities of the system are given in 
Table~\ref{table4}. The semi-major axis of the binary system is calculated to 
be $a = 1.906 ~R_{\odot}$. 

\begin{figure*}
\vspace{0.02\linewidth}
\begin{tabular}{cc}
\vspace{+0.03\linewidth}
\resizebox{0.50\linewidth}{!}{\includegraphics*{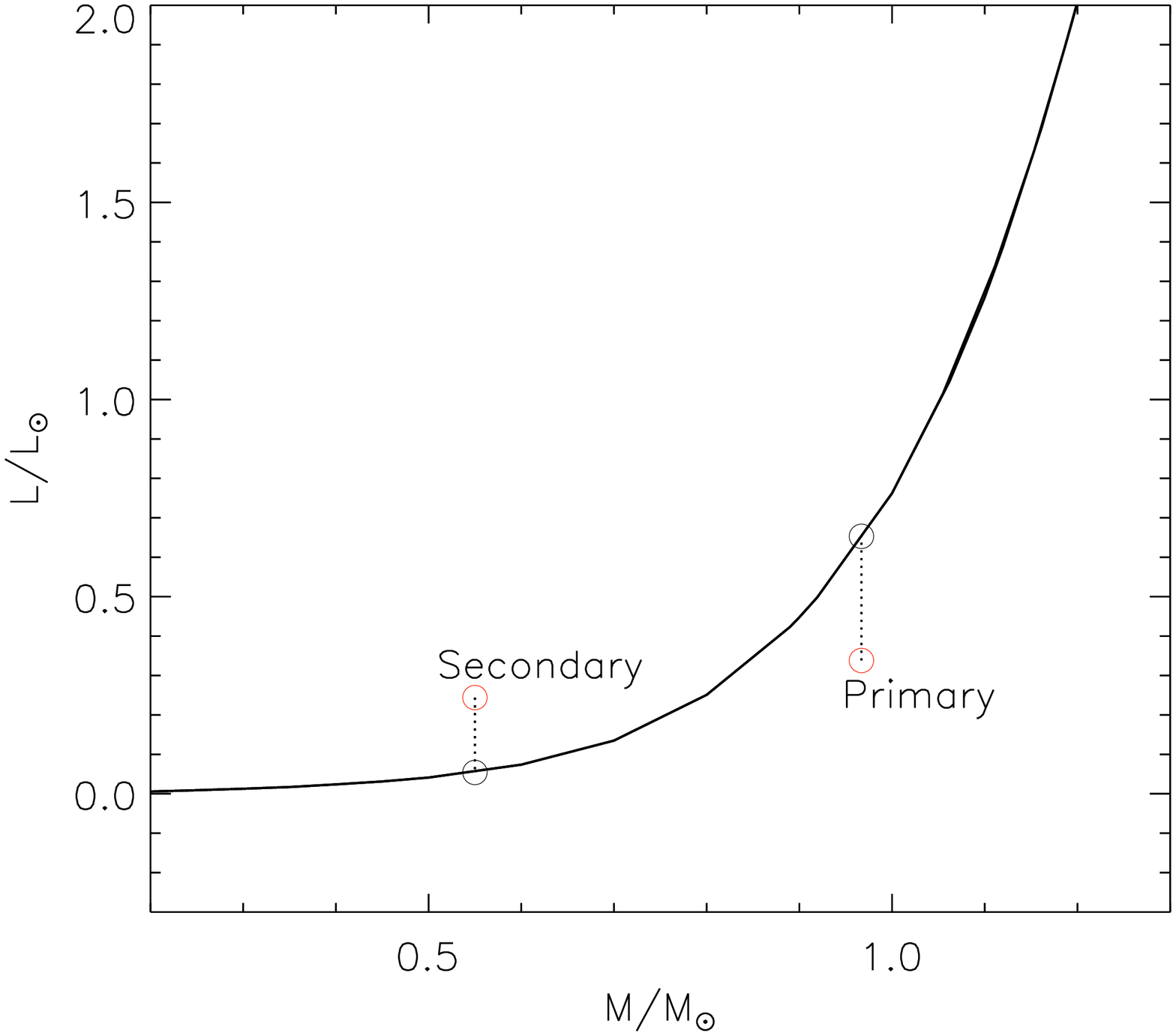}}
\resizebox{0.50\linewidth}{!}{\includegraphics*{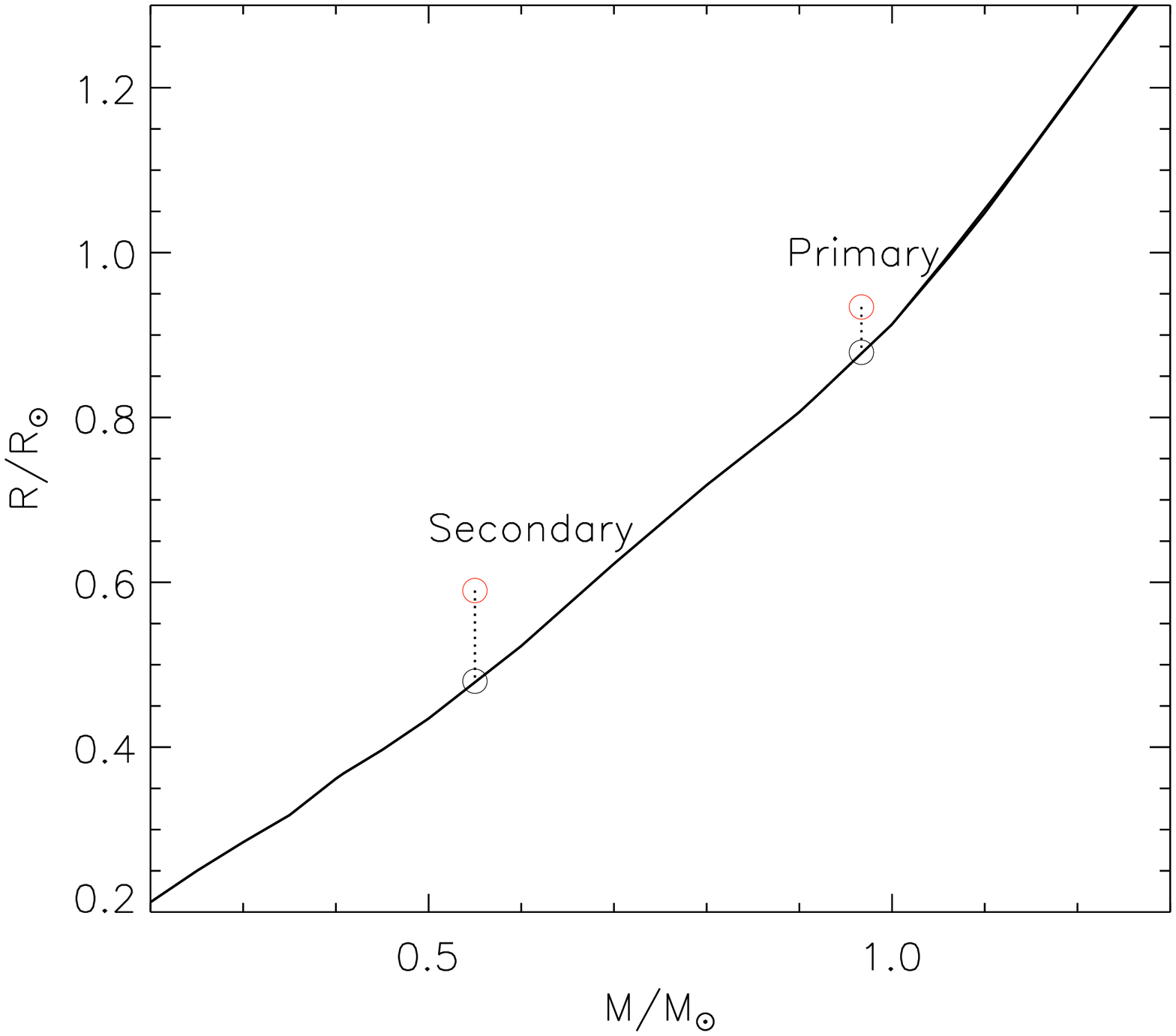}}
\vspace{-0.04\linewidth}
\end{tabular}
\caption{Locations of the components of EK Com on the main sequence 
mass-luminosity and mass-radius diagrams. The solid line represents 1 Gyr 
isochrones from \citet{gir2000} with solar chemical composition.}
\label{mass_radius}
\end{figure*}

Fig.~\ref{mass_radius} shows the location of the components of EK Com on 
the theoretical mass-luminosity (M-L) and mass-radius (M-R) diagrams along 
with the 1 Gyr theoretical isochrones from \citet{gir2000} for the population 
I stars with the solar composition $(X,Y,Z) = 0.708, 0.273, 0.019$. 

According to \citet{moc1981}, the densities $\bar 
\rho_{1},\,\bar \rho_{2}$ for the primary and secondary components are 
respectively described by 
\begin{eqnarray}
\bar \rho_{1} = \frac{0.0189}{r_{1}^{3}(1+q)P^{\,2}}, \bar \rho_{2} = 
\frac{0.0189q}{r_{2}^{3}(1+q)P^{\,2}}  
\end{eqnarray}
We find mean densities for the primary and secondary component to be 
1.672~$\rho_{\odot}$, 2.323~$\rho_{\odot}$ respectively. 
\begin{table}
\begin{center}
\caption{Absolute parameters of EK Com}
\label{table4}
\begin{tabular}{lcc}
\hline \hline
Parameter                & Primary   & Secondary  \\ \hline
Mass ($M_{\odot}$)       & 0.967     & 0.338      \\ \hline
Radius ($R_{\odot}$)     & 0.934     & 0.590      \\ \hline
Luminosity ($L_{\odot}$) & 0.550     & 0.243      \\ \hline 
\end{tabular}
\end{center}
\end{table}

We have also calculated the following parameters as described in \citet{csi2004}: the bolometric luminosity ratio ($\lambda$), energy transfer 
parameters ($\beta$), $\beta_{\rm excess}$, $\beta_{\rm envelope}$, 
$\beta_{\rm corr}$, amount of the transferred luminosity ($\Delta L$). The 
values of the obtained parameters are given in Table~\ref{table5}. 
\begin{table}
\begin{center}
\caption{bolometric luminosity ratio, energy transfer parameters}
\label{table5}
\begin{tabular}{cccccc}
\hline \hline
$\lambda$ & $\beta$ & $\beta_{\rm excess}$ & $\beta_{\rm envelope}$ & $\beta_{\rm corr}$ & $\Delta L$ \\ \hline
0.417 & 0.711 & 0.005 & 0.706 & 0.704 & 0.159 \\ \hline 
\end{tabular}
\end{center}
\end{table}
\section{Distance determination to EK Com}
The standard $V$ magnitude of the comparison star is 12.060 (SA96).~We now 
find the differential magnitude ($\Delta V$) between the variable and the 
comparison star at phases 0, 0.25, 0.50 and 0.75 respectively. From the spot
fitted light curve, we find that $\Delta V = 0.6594~(\Phi = 0.0),\,
0.0095~(\Phi = 0.25),\,0.6451~(\Phi = 0.50),\,0.0462~(\Phi = 0.75)$.~A brief magnitude calibration gives the standard magnitude of 
EK Com $V = 12.719~(\Phi = 0.0), 12.069~(\Phi = 0.25),\,
12.705~(\Phi = 0.50),\,12.106~(\Phi = 0.75)$.~With the derived luminosities 
for both the components and the bolometric correction values found from the 
temperatures of the components following \citet{flo1996}, we obtain the 
absolute magnitudes of the components as 5.652 mag and 6.476 mag respectively.
For the primary and secondary components, the bolometric correction values 
used are $-0.253$ and  $-0.190$ respectively. The combined absolute 
magnitude of the system is then computed as $M_{\rm V} = 5.235$ mag. On 
the other hand, \citet{ruc1997b} derived the following empirical relation 
with an accuracy of $\pm0.1$ to calculate the absolute magnitude of W UMa 
systems.
\begin{equation}
M_{V} = -4.44\,\log\,P+3.02(B-V)_{0}+0.12.
\label{eqbol}
\end{equation}
Using the above equation, the absolute magnitude of EK Com can be calculated as
5.000 mag.~Our calculated value of the absolute magnitude of EK Com 5.235 mag 
is nearly 0.23 mag fainter than the value obtained from Eq.~\ref{eqbol}. 
We now take the visual magnitude $V = 12.106$ at phase $\Phi = 0.75$. The 
interstellar absorption is taken as $A_{V} = 0.031$. The distance modulus of 
EK Com is then computed to be $(m-M) = 6.840$ mag using our computed value 
of the absolute magnitude of 5.235 mag for the binary system. This corresponds 
to a distance of $233.346 \rm ~pc$ or parallax of $4.30 \rm ~ mas$. On the 
other hand, if the value of the absolute magnitude of 5.000 mag obtained from
Eq.~\ref{eqbol} is used, the distance of EK Com can be calculated as $260.016$ 
pc or a parallax of $3.8$ mas.    
\section{Summary \& Conclusions}
We have presented a photometric solution for the eclipsing binary EK Com 
obtained from the new high-precision time-series CCD photometric observations 
with complete phase coverage in the V band. The new observations indicate that 
the star may be of A-type with a flat secondary minimum. By using the 
WD code, we have found that the system is a total-eclipsing 
binary star with an orbital inclination of $i[^{o}] = 89.800 \pm 0.075$ and a
high degree of over-contact configuration 33.0\%. The difference in the mean 
temperatures between the two components is found to be $141 \pm 10~K$. 
The observed light curve is quite asymmetric, especially at quadrature levels, 
because of surface activity on either of the two components. This feature of 
the light curve is explained assuming a hot spot on the  primary component. 

It has been found that the photometric mass ratio determined from the WD light 
curve modelling approach is reliable for a totally eclipsing over-contact 
binary. Keeping in view of this fact, the absolute physical parameters of the 
two components are determined based on the results of the light curve solution 
and by the MV96 method. The luminosity of the star is calculated using the 
temperature and radius of each of the individual components. The photometric 
spot solutions given in Table~\ref{table3} are tentative, as it has been 
observed that with the spot solutions included in the WD code, the whole 
light curve can be easily fitted, but a serious problem of uniqueness of the 
solution persists \citep{mac1993} unless other means of investigation such 
as Doppler Imaging techniques are applied \citep{mac1994}. 

The spectroscopic radial velocity observations and long term photometric monitoring of the 
star is important to determine its actual type and to answer whether the 
deeper rounded shape of the primary eclipse seen in our observations is either 
due to the actual configuration of the system or due to its higher level of 
spot magnetic activity at different phases or both.
   
\section*{Acknowledgments}
The authors thank $IUCAA$ for providing telescope time available on the IGO 
2-m telescope.  The authors thank the anonymous referee for useful comments 
and suggestions which improved the presentation of the paper. The authors 
thank the staff at IGO for their active support during the course of 
observations. SD thanks CSIR, Govt. of India for a Senior Research Fellowship. 
The use of the SIMBAD, ADS, ESO DSS databases is gratefully acknowledged. This publication makes use of data products from the Two Micron All Sky Survey, which is a joint project of the University of Massachusetts and the Infrared Processing and Analysis Center/California Institute of Technology, funded by the National Aeronautics and Space Administration and the National Science Foundation. ESO-MIDAS was used as a part of the data analysis. 

\bibliographystyle{elsarticle-harv}
\bibliography{paper}
\end{document}